\documentclass[prd,preprint,tightenlines,floatfix,showpacs,preprintnumbers,nofootinbib,eqsecnum]{revtex4}

\usepackage[dvips,final]{graphicx}
  \usepackage{amssymb}
   \usepackage{amsmath}
    \usepackage{amsfonts}
     \usepackage{epsfig}
      \usepackage{bm}

\renewcommand{\epsilon}{\varepsilon}

\begin{document}

\thispagestyle{empty} \preprint{\hbox{}} \vspace*{-10mm}

\title{Diagonalization of the neutralino mass matrix
and boson-neutralino interaction}
\author{V.~A.~Beylin}

\email{vbey@rambler.ru}

\author{V.~I.~Kuksa}

\email{kuksa@list.ru}

\author{G.~M.~Vereshkov}

\email{gveresh@gmail.com}

\affiliation{Institute of Physics,
Southern Federal University ,
Rostov-on-Don 344090, Russia}

\author{R.~S.~Pasechnik}

\email{rpasech@theor.jinr.ru}

\affiliation{Bogoliubov Laboratory of Theoretical Physics, JINR,
Dubna 141980, Russia}

\date{\today}

\begin{abstract}
We analyze a connection between the neutralino mass sign, parity and structure of the neutralino-boson interaction. Correct calculation of spin-dependent and spin-independent contributions to neutralino-nuclear scattering should consider this connection. A convenient diagonalization procedure, based on the exponential parametrization of unitary matrix, is suggested.
\end{abstract}


\maketitle

\section{Introduction}

Superpartners of gauge and Higgs fields play an important role in
SUSY phenomenology. In particular, neutralino dark matter (DM)
in the SUSY framework was considered in detail (see, for example, Refs. \cite{1}-\cite{7}),
and these investigations were adapted to astrophysics. So, an analysis
of the neutralino system and the structure of gauge bosons
interaction with the neutralino and chargino is important for the DM
description and the study of astrophysical data.

In many phenomenological works both the neutralino mass spectrum
and the structure of states follow from the formal
diagonalization of the neutralino mass form by an orthogonal (real) matrix
\cite{1}, \cite{8} - \cite{11}. Such a procedure does not consider some important features of
the structure of the Majorana states, related with the sign of the mass.
These features are connected with the structure of the neutralino-boson interactions which, in turn, defines the
peculiarity of the neutralino-nucleon scattering.

The most complete and comprehensive analysis of the neutralino
system has been performed in \cite{12} - \cite{13}. In
these papers, the diagonalization of the neutralino mass matrix is
considered in detail in the MSSM and some of its extensions. Special
attention was paid to the building of neutralino states with positive
masses. However, due to the complexity of the general diagonalization
formalism it is difficult to trace a link between the sign of mass
and the structure of the neutralino-nucleon interactions.

In this paper, we analyze the features of the neutralino structure and
interactions which are directly related with the sign of mass. We
consider the simplest case when this connection is transparent
and convenient for illustration. In the second section we compare two ways of the diagonalization -- by orthogonal and
unitary diagonalyzing matrices. These two variants lead to neutralino states with opposite and equal signs of masses.
They are formally equivalent and related by a field redefinition (see Section 2).
But the negative mass of the neutralino (as it occurs for the first case), has to be
taken into consideration in the consequent calculations. Disregarding this important feature, it is possible to get an incorrect conclusion on the spin-dependent (SD)
and spin-independent (SI) contributions into the neutralino-nuclear
cross-section \cite{10}.

Redefinition of the field reveals a link between the sign of the mass and transformation
properties of Majorana spinors with respect to inversion (i.e.
parity). An analogous connection between the sign of the mass and parity
was revealed for the case of massive Majorana neutrino in
Refs.~\cite{wolf} and \cite{bilenky}. In Section 3, we present the compact Lagrangian of the boson-neutralino-chargino interaction in terms of a redefined field, which is convenient for phenomenological applications.

In the fourth section, we consider the neutralino mass matrix
diagonalization by means of a unitary matrix giving all
positive masses. Thus, the standard calculation rules can be
kept unchanged, and there is no need to check the sign of the mass or
redefine the field. A convenient diagonalization procedure based
on the exponential parametrization of the unitary matrix is discussed.
This procedure is formalized in a perturbative calculation scheme analogous to \cite{12}.
However, our scheme needs a smaller number of input
parameters and gives all expressions in a quite compact form, which is useful
for calculations. The method suggested is generalized for the case of mass matrix with complex parameters (Appendix B).

\section{Neutralino parity and structure of boson-neutralino interaction}

Now we analyze the connection between the signs of the neutralino mass and the structures of the neutralino-bozon interaction when $M_Z/M_k \rightarrow 0$ and $M_k$ is $M_1, M_2$ or $\mu$. In this limit, the analysis is simplified considerably, but the results can be used in the general case too.
This limit is approximately realized in Split SUSY scenarios
\cite{3} and strictly takes place at high temperatures $T\gg
E_{EW}$, when the Higgs condensate is melted (the high symmetry phase).
For completeness, we give the well-known minimal formalism that we need in the following analysis.

If the mixing of gauge and Higgs fermions is neglected, the mass
term of higgsino-like Majorana fields has the Dirac form \cite{Cheng}:
\begin{equation}\label{E:2.1}
 M_h=\frac{1}{2}\mu(\bar{H}^0_{1R}H^0_{2L}+\bar{H}^0_{2R}H^0_{1L})
 +h.c.\,\,.
\end{equation}
This form can be represented by a ($2\times2$)\,- mass matrix which is known as the specific matrix with zero trace:
\begin{equation}\label{E:2.2}
    \mathbf{M}_2=
   \begin{pmatrix}
     0&\mu\\
     \mu&0
   \end{pmatrix}.
\end{equation}
There are two ways to diagonalize this matrix.
The formal procedure using the orthogonal matrix
$\mathbf{O}_2$ leads to a spectrum with opposite signs:
\begin{equation}\label{E:2.3}
 \mathbf{O}^T_2\mathbf{M}_2\mathbf{O}_2=\begin{pmatrix}
 \mu&0\\
 0&-\mu \end{pmatrix},\,\,\,\mathbf{O}_2=\frac{1}{\sqrt{2}}\begin{pmatrix}
 1&\,\,1\\
 1&-1 \end{pmatrix}\,,\,\,\,m_a=(\mu,-\mu)\,,
\end{equation}
where
$\mathrm{Tr}\{\mathbf{O}^T_2\mathbf{M}_2\mathbf{O}_2\}=\mathrm{Tr}\{\mathbf{M}_2\}=0$ (trace conservation).
In this case, one of the Majorana fields has a negative mass, regardless of the sign of $\mu$.
The matrix $\mathbf{M}_2$ can also be diagonalized by the unitary
complex matrix $\mathbf{U}_2$, giving masses with the same sign:
\begin{equation}\label{E:2.4}
 \mathbf{U}^T_2\mathbf{M}_2\mathbf{U}_2=\begin{pmatrix}
 \mu&0\\0&\mu\end{pmatrix},\,\,\,\mathbf{U}_2=\frac{1}{\sqrt{2}}
 \begin{pmatrix}1\,\,&i\\1&-i\end{pmatrix}\,,\,\,\,m_a=(\mu,\mu).
\end{equation}
The diagonalization (\ref{E:2.4}) is equivalent
to the procedure (\ref{E:2.3}) with the redefinition
$\chi\rightarrow i\gamma_5 \chi$ of the non-chiral (full) field with $m=-\mu$. The last transformation is equivalent to $\chi_{R,L}\to \pm i \chi_{R,L}$ for the chiral components. Note that there is an infinite set of unitary matrices
$\mathbf{U}_{\phi}=\mathbf{U}_2\cdot\mathbf{O}_\phi$ which
diagonalize the mass matrix $\mathbf{M}_2$ (see also
\cite{12c}, Appendix A.2):
\begin{equation}\label{E:2.5}
 \mathbf{U}_{\phi}=\frac{1}{\sqrt{2}}\begin{pmatrix}e^{i\phi}&ie^{i\phi}\\
 e^{-i\phi}&-ie^{-i\phi}\end{pmatrix},\,\,\,\mathbf{O}_{\phi}=\begin{pmatrix}
 \cos \phi&-\sin \phi\\ \sin \phi&\cos \phi\end{pmatrix}.
\end{equation}
The additional
$O_2$\,-\,symmetry (see Appendix B) leads to a free parameter arising in the general case.

Dealing with the spinor field we should
take into account the sign of its mass in the propagator and
polarization matrix or redefine the field with a negative mass. As a rule this feature is not considered in phenomenological applications (see, for example, Refs. \cite{8}-\cite{11}). From the redefinition $\chi'=i\gamma_5\chi$, it follows that the transformation (relative to inversion) properties of Majorana fields having opposite mass signs are different. As a result, we have one usual Majorana field and one pseudo-Majorana field.

The gaugino mass subform is of the standard Majorana type \cite{Cheng} and has no specific features. The signs of the masses for $\chi_1$ and $\chi_2$ are defined by the signs of $M_1$ and $M_2$ in the case of small mixing. They can be made positive by a redefinition. Note that the redefinition procedure always influences the mixing terms of the mass matrix and should be taken into account in the general case (see Sect.4).

Now we consider the connection between the structure of
the boson-neutralino interaction and the relative sign of the neutralino masses.
For simplicity, we show this connection in the pure higgsino
approximation. The contribution of terms caused by mixing is considered
in the next section. We here present a short comparative analysis of
the calculation rules in two cases: when the masses of $\chi_3$ and $\chi_4$
have different signs (diagonalization (\ref{E:2.3})) and when they have the same
signs (diagonalization (\ref{E:2.4})). The initial Lagrangian is
\begin{equation}\label{E:3.1}
 L_{int}=\frac{1}{2}g_Z
 Z_{\mu}(\bar{H}^0_{1L}\gamma^{\mu}H^0_{1L}+\bar{H}^0_{2R}\gamma^{\mu}H^0_{2R})\,,
\end{equation}
where $g_Z=g_2/\cos \theta_W$. The diagonalizations (\ref{E:2.3}) and (\ref{E:2.4}) lead to the following forms of neutralino-boson interactions, respectively:
\begin{equation}\label{E:3.2}
 (1)\; L_{int}=-\frac{1}{2}g_Z
 Z_{\mu}\bar{\chi}_3\gamma^\mu\gamma_5\chi^{'}_4;\qquad\quad
 (2)\; L_{int}=\frac{i}{2}g_Z Z_{\mu}\bar{\chi}_3\gamma^{\mu}\chi_4.
\end{equation}
In Eqs.~(\ref{E:3.2}) the first case with opposite signs
$(\mu,-\mu)$ can be transformed into the second case with the same
signs $(\mu,\mu)$ by the redefinition $i\gamma_5\chi^{'}_4=\chi_4$. Here
we show that both Lagrangians in Eqs.~(\ref{E:3.2})
lead to the same result without any field redefinition if the
negative sign of $\chi^{'}_4$ mass is considered in
the calculations. In other words, both structures in
Eqs.~(\ref{E:3.2}) lead to the parity-conserving vector interaction
which gives the spin-independent contribution to the
neutralino-nucleon scattering \cite{Kur}.

Let us consider, for example, the process of the scattering $\chi_3
q\rightarrow \chi_4 q$ with $t$-channel exchange of a $Z$-boson ($t^2\ll
M^2_Z$). In both cases, the amplitudes of this process are
\begin{align}\label{E:3.3}
 (1)\quad \mathcal{M}_1&\sim\bar{\chi}^{'+}_4(p_2)\gamma^{\mu}\gamma_5 \chi^-_3(p_1)\cdot
 \bar{q}^+(k_2)\gamma_{\mu}(c_q-\gamma_5)q^-(k_1)\,,\notag\\
 (2)\quad \mathcal{M}_2&\sim\bar{\chi}^+_4(p_2)\gamma^{\mu}\chi^-_3(p_1)\cdot\bar{q}^+(k_2)\gamma_{\mu}
 (c_q-\gamma_5)q^-(k_1)\,.
\end{align}
Formally, the amplitudes $\mathcal{M}_1$ and $\mathcal{M}_2$ have a different structure. Therefore, one can drow a wrong conclusion about the contributions to the
spin-dependent and spin-independent parts of the cross-section , if the negative sign of $\chi^{'}_4$ mass has not been taken into account. However, taking into account the negative sign of $\chi^{'}_4$ in
the polarization matrix allows one to get the same result for both
cases. If $\chi$ is in an initial or final state, the polarization
matrix of the field $\chi$ in $\mathcal{M}^{+}\mathcal{M}$ is defined by (for positive mass $m_{\chi}=\mu>0$)
\begin{equation}\label{E:3.4}
 \sum_{\sigma}\chi^{\mp}_{\sigma}(p)\bar{\chi}^{\pm}_{\sigma}(p)=\frac{1}{2p^0}(\hat{p}\pm
 \mu),
\end{equation}
or (for negative mass $m_{\chi^{'}}=-\mu$)
\begin{equation}\label{E:3.5}
 \sum_{\sigma}\chi^{'\mp}_{\sigma}(p)\bar{\chi}^{'\pm}_{\sigma}(p)=\frac{1}{2p^0}(\hat{p}\mp
 \mu).
\end{equation}
With the help of Eq.~(\ref{E:3.5}), we get
\begin{equation}\label{E:3.6}
 \mathcal{M}^+_1\mathcal{M}_1\sim
 \mathrm{Tr}\{(\hat{p}_2-\mu)\gamma^{\mu}\gamma_5(\hat{p}_1+\mu)\gamma^{\nu}\gamma_5\}=
 \mathrm{Tr}\{(\hat{p}_2+\mu)\gamma^{\mu}(\hat{p}_1+\mu)\gamma^{\nu}\}.
\end{equation}
One can get the same expression for $\mathcal{M}^+_2 \mathcal{M}_2$ using the standard definition (\ref{E:3.4}) of the polarization matrix. This feature
should be included in an analysis of neutralino-nucleon
scattering. From the interaction Lagrangian only, without consideration of the mass
signs, we cannot drow any valuable conclusions on the
SD or SI contributions. In particular, the
bilinear structures $\bar{\chi}_3\gamma_{\mu}\chi_4$ and
$\bar{\chi}_3\gamma^{\mu}\gamma_5\chi^{'}_4$ are vectors, while
$\bar{\chi}_3\gamma_{\mu}\chi^{'}_4$ and
$\bar{\chi}_3\gamma^{\mu}\gamma_5\chi_4$ are axial vectors.
Analogously, $\bar{\chi}_3\chi^{'}_4$ and $\bar{\chi}_3\gamma_5\chi_4$ are
pseudoscalars, while $\bar{\chi}_3\gamma_5\chi^{'}_4$ and $\bar{\chi}_3\chi_4$ are scalars. Thus, the analysis of the
neutralino-nucleon interaction has to take into account neutralino
transformation properties. As a rule, in the bulk of papers this
feature has not been considered explicitly and mistaken conclusions
can be obtained in calculations of the SD and SI contribution to
the neutralino-nucleon interaction. In particular, for the current
structure $\bar{\chi}_i\gamma^{\mu}\gamma_5\chi_k Z_{\mu}$ it is
possible to obtain SD or SI neutralino-nucleon cross sections
depending on the neutralino relative parity. For instance, in
Refs.~\cite{10}, \cite{14}, \cite{15}] the same
current structure was considered without any comments on this
important peculiarity. From our analysis, it follows that in the
case discussed, the neutralino-boson interaction gives the main
contribution to the spin-independent part of the cross-section \cite{Kur}.

An analogous feature is in order when $\chi^{'}_4$ is in an
intermediate state, for example, in the process $\chi_3 Z\rightarrow
\chi^{'}_4\rightarrow \chi_3 Z$. The amplitude $M_1$ of the process
is
\begin{equation}\label{E:3.7}
 \mathcal{M}_1\sim
 \bar{\chi}^+_3(p_2)\gamma^{\mu}\gamma_5(\hat{q}-\mu)\gamma^{\nu}\gamma_5\chi^-_3(p_1)
 e^Z_{\mu}e^Z_{\nu}=\bar{\chi}^+_3(p_2)\gamma^{\mu}(\hat{q}+m_{\chi})\gamma^{\nu}\chi^-_3(p_1)
 e^Z_{\mu}e^Z_{\nu}.
\end{equation}
In Eq.~(\ref{E:3.7}) we use the propagator $\sim(\hat{q}-\mu)$ for
the field $\chi^{'}_4$ with negative mass $m_{\chi^{'}}=-\mu$,
whereas the standard propagator is $\sim (\hat{q}+\mu)$. So, the mass sign been taking in account leads to the same result for the amplitudes $\mathcal{M}_1$ and $\mathcal{M}_2$, where $\mathcal{M}_2$ describes the same process with redefined $\chi_4$ in an intermediate state.

\section{Gauge boson-neutralino-chargino interactions}

In this section, we give compact expressions for the Lagrangian of gauge bozon-neutralino-chargino interactions in the case of small mixing. These expressions are convenient for calculations in cosmology.
This Lagrangian follows from Eqs. (\ref{E:7.14}) - (\ref{E:7.16}) (Appendix A) as a result of the shift:
\begin{eqnarray}
L_{int}&=&\frac{i}{2}g_2\epsilon_{abc}\bar{W}^a\gamma^{\mu}W^cW^b_{\mu}
         -\frac{1}{2}g_1\bar{H}^-_1\gamma^{\mu}H^-_{1L}B_{\mu}+\nonumber\\
       &+&\frac{1}{2}g_1\bar{H}^+_2\gamma^{\mu}H^+_{2L}B_{\mu}+\nonumber\\
       &+&\frac{1}{\sqrt{2}}g_2W^+_{\mu}(\bar{H}^0_1\gamma^{\mu}H^-_{1L}+\bar{H}^+_2\gamma^{\mu}H^0_{2L})+\nonumber\\
       &+&\frac{1}{\sqrt{2}}g_2W^-_{\mu}(\bar{H}^-_1\gamma_{\mu}H^0_{1L}+\bar{H}^0_2\gamma^{\mu}H^+_{2L})+\label{E:4.1}\\
       &+&\frac{1}{2}g_2W^3_{\mu}(-\bar{H}^-_1\gamma^{\mu}H^-_{1L}+\bar{H}^+_2\gamma^{\mu}H^+_{2L}+
       \bar{H}^0_1\gamma^{\mu}H^0_{1L}-\nonumber\\
       &-&\bar{H}^0_2\gamma^{\mu}H^0_{2L})-\frac{1}{2}g_1\bar{H}^0_1\gamma^{\mu}H^0_{1L}B_{\mu}
         +\frac{1}{2}g_1\bar{H}^0_2\gamma^{\mu}H^0_{2L}B_{\mu}.\nonumber
\end{eqnarray}

Let us consider the case $M_Z\ll\mu, M_{1,2}$, which can be used in
Split SUSY models \cite{2}, \cite{3}, \cite{4}, \cite{10}. The physical states of the neutralino in the zeroth order of the mixing were defined in Section 2, and the chargino states in Appendix B:
\begin{eqnarray}\nonumber
 &&{}\chi_1=W^3,\quad\chi_2=B,\quad\chi_3=(H^0_1+H^0_2)/\sqrt{2},\\
 &&{}\chi_4=i\gamma_5(H^0_1-H^0_2)/\sqrt{2};\label{E:4.2}\\
 &&{}\tilde{H}=-i\gamma_5(H^-_{1L}+H^+_{2R}),\quad \tilde{W}=(W_1+iW_2)/\sqrt{2}.
 \nonumber
\end{eqnarray}
In Eqs. (\ref{E:4.2}) we do not use charge sign notation for the Dirac
fields $\tilde{H}$ and $\tilde{W}$ (in contrast to $W^{\pm}_{\mu}$) in analogy to the Standard Model notation. From the structure of $\tilde{H}$ in (\ref{E:4.2}), it follows that the components $H^-_{1L}$ and $H^+_{2R}$ correspond to particle and anti-particle parts in a Weyl basis. Using the definitions (\ref{E:4.2}) we represent $L_{int}$ in the form
\begin{align}\label{E:4.3}
 L_{int}&=g_2W^+_{\mu}(\bar{\chi}_1\gamma^{\mu}\tilde{W}-\frac{i}{2}\bar{\chi}_3\gamma^{\mu}\tilde{H}
          -\frac{1}{2}\bar{\chi}_4\gamma^{\mu}\tilde{H})\notag\\
        &+g_2W^-_{\mu}(\bar{\tilde{W}}\gamma^{\mu}\chi_1+\frac{i}{2}\bar{\tilde{H}}\gamma^{\mu}\chi_3
          -\frac{1}{2}\bar{\tilde{H}}\gamma^{\mu}\chi_4)\notag\\
        &-g_2\cos{\theta_W}Z_{\mu}\bar{\tilde{W}}\gamma^{\mu}\tilde{W}-\frac{g_2}{2\cos{\theta_W}}
          \cos{2\theta_W}Z_{\mu}\bar{\tilde{H}}\gamma^{\mu}\tilde{H}\\
        &+\frac{ig_2}{2\cos{\theta_W}}Z_{\mu}\bar{\chi}_3\gamma^{\mu}\chi_4
          -eA_{\mu}\bar{\tilde{W}}\gamma^{\mu}\tilde{W}-eA_{\mu}\bar{\tilde{H}}\gamma^{\mu}\tilde{H}.\notag  \end{align}

The first order corrections to the $Z\chi_i\chi_k$ interaction caused
by the mixing (see Appendix B) are
\begin{align}\label{E:4.4}
 L^{(1)}_{mix}=&\frac{g_2}{2\cos{\theta_W}}Z_{\mu}(-\frac{im_2}{M_1-\mu}\bar{\chi}_1\gamma^{\mu}\chi_3
           +\frac{im_4}{M_2-\mu}\bar{\chi}_2\gamma^{\mu}\chi_3\notag\\
           &-\frac{m_1}{M_1+\mu}\bar{\chi}_1\gamma^{\mu}\gamma_5\chi_4
           +\frac{m_3}{M_2+\mu}\bar{\chi}_2\gamma^{\mu}\gamma_5\chi_4),
\end{align}
where the $m_k$ are defined by Eq. (\ref{E:5.5}) in the next section.
From Eq. (\ref{E:4.4}), one can see that the interactions of $\chi_3$
and $\chi_4$ with $\chi_{1,2}$ have a different structure. This effect
is directly connected with different signs of the masses of
the non-redefined fields. Note also that the bino-like
neutralino $\chi_2\approx B$ does not interact with gauge bosons in
the zero mixing approximation, but it interacts with the scalar Higgs
field and $\chi_{3,4}$.

In the pure higgsino limit, $\chi_3$ and $\chi_4$ constitute the neutral
Dirac field $\tilde{H}^0=(\chi_3+i\chi_4)/\sqrt{2}$ and the part of Eq.
(\ref{E:4.3}) can be represented in the form (here we omit the heavy
states $\tilde{W}$ and $\chi_2$)
\begin{eqnarray}\nonumber
 L^D_{int}&=&-\frac{ig_2}{2}W^+_{\mu}\bar{\tilde{H}}^0\gamma^{\mu}\tilde{H}+
           \frac{ig_2}{2}W^-_{\mu}\bar{\tilde{H}}\gamma^{\mu}\tilde{H}^0+\\
          &+&\frac{g_2}{2\cos{\theta_W}}Z_{\mu}\bar{\tilde{H}}^0\gamma_{\mu}\tilde{H}^0-\label{E:4.5}\\
          &-&\frac{g_2}{2\cos{\theta_W}}\cos{2\theta_W}Z_{\mu}\bar{\tilde{H}}\gamma_{\mu}\tilde{H}-
           eA_{\mu}\bar{\tilde{H}}\gamma^{\mu}\tilde{H}\,.\nonumber
\end{eqnarray}
The Dirac representation (\ref{E:4.5}) of the
boson-neutralino-chargino interactions involving a small
mixing of the Higgs fermion with the gauge ones is formal (unphysical)
but is convenient for our calculations. In this case, we avoid some
complications of the Feynman rules, caused by the Majorana nature of $\chi_3$
and $\chi_4$ \cite{16} - \cite{18}. By direct calculation we have
checked that both ways lead to the same results for
the annihilation and co-annihilation cross-sections \cite{19}.

\section{Diagonalization of the neutralino mass matrix by unitary matrix
 with exponential parametrization}

In this section, we consider diagonalization
of the $4\times4$ mass matrix with real parameters $\mu, M_1, M_2$.  Generalization of the approach for a matrix with complex parameters is considered in Appendix B. As follows from Section 2, the sign $\mu$ is not essential, and $M_{1,2}$ can be made positive by a redefinition of the gauge fermion. The neutral fermion mass form follows from the SUSY Lagrangian (Eqs.~(\ref{E:7.15}) and (\ref{E:7.16}) in Appendix A) after the shift
\begin{equation}\label{E:5.1}
 L_m=-\frac{1}{2}(\bar{\phi}_R)^T\mathbf{M}_0\phi_L+h.c.,
\end{equation}
where $(\phi)^T=(B,W^3,H^0_1,H^0_2)$ and
\begin{equation}\label{E:5.2}
 \mathbf{M}_0=\begin{pmatrix}M_1&0&-iM_Zs_{\theta}c_{\beta}&iM_Zs_{\theta}s_{\beta}\\
 0&M_2&iM_Zc_{\theta}c_{\beta}&-iM_Zc_{\theta}s_{\beta}\\
 -iM_Zs_{\theta}c_{\beta}&iM_Zc_{\theta}c_{\beta}&0&\mu\\
 iM_Zs_{\theta}s_{\beta}&-iM_Zc_{\theta}s_{\beta}&\mu&0
 \end{pmatrix},
\end{equation}
where $s_{\theta}=\sin{\theta}$ and $c_{\beta}=\cos{\beta}$. The matrix (\ref{E:5.2}) differs from the commonly
used one by the presence of the imaginary unit in the mixing terms. One can
go to a real traditional matrix $M^{'}_0$ by a redefinition
$H^{'}_a=i\gamma_5H_a$, where $a=1,2$. As a result we get the
standard matrix following from (\ref{E:5.2})
under the formal transition $iM_Z\rightarrow M_Z$ and
$\mu\rightarrow-\mu$. However, implying our diagonalization
procedure there is no need to do this transformation (see also Ref.
\cite{15}).

It is convenient to analyze the diagonalization of the matrix
(\ref{E:5.2}) with the help of the intermediate transformation
\begin{equation}\label{E:5.3}
 \mathbf{M}_I=\mathbf{U}^T_I\mathbf{M}_0\mathbf{U}_I;\,\,\,
 \mathbf{U}_I=\begin{pmatrix}\mathbf{1}&\mathbf{0}\\
 \mathbf{0}&\mathbf{U}_2\end{pmatrix}.
\end{equation}
Here $\mathbf{1}$ and $\mathbf{0}$ are the identity and zero
$(2\times2)$\,-\,matrices, and $\mathbf{U}_2$ is defined
by Eq. (\ref{E:2.4}) in the pure higgsino limit. Then the
intermediate mass matrix has the form
\begin{equation}\label{E:5.4}
 \mathbf{M}_I=\begin{pmatrix}
 M_1&0&-im_1&m_2\\0&M_2&im_3&-m_4\\-im_1&im_3&\mu&0\\m_2&-m_4&0&\mu
 \end{pmatrix},
\end{equation}
where
\begin{eqnarray}
m_1&=&M_Z \sin \theta_W(\cos \beta-\sin \beta)/\sqrt{2},\nonumber\\
m_2&=&M_Z \sin \theta_W(\cos \beta+\sin \beta)/\sqrt{2},\nonumber\\
m_3&=&M_Z \cos \theta_W(\cos \beta-\sin \beta)/\sqrt{2},\label{E:5.5}\\
m_4&=&M_Z\cos \theta_W(\cos \beta+\sin \beta)/\sqrt{2}.\nonumber
\end{eqnarray}
Intermediate fields are defined by $\phi_I=\mathbf{U}_I\phi$; that is,
$(\phi_I)^T=(B,W^3,\chi^I_3,\chi^I_4)$, where $\chi^I_3$ and
$\chi^I_4$ are defined by Eq.~(3.2).
The use of the intermediate mass matrix provides the positivity of
the higgsino-like neutralino masses and leads to the "quasidiagonal"
structure of the matrix in the case of small mixing.

The matrix (\ref{E:5.4}) is symmetric and complex, but it is not Hermitian.
The spectrum of $\mathbf{M}_I$ is real and has a simple form.
However, it is not a mass spectrum of the neutralino, because the
diagonalization of the neutralino mass matrix
$\mathbf{U}^T_I\mathbf{M}_I\mathbf{U}_I=diag(m_k)$ differs from the
one defined by $\mathbf{U}^+\mathbf{M}_I\mathbf{U}=diag(\lambda_k)$.
In the last case, $\mathbf{U}$ is built of the eigenvectors of $\mathbf{M}_I$ and
the $\lambda_k$ are the eigenvalues of $\mathbf{M}_I$. According to the
theorem 4.4.4 (Takagi expansion) from Ref. \cite{20}, any complex
symmetric matrix can be diagonalized by the unitary matrix
$\mathbf{U}$:
\begin{equation}\label{E:5.6}
 \mathbf{U^+MU^*}=diag(m_k),\,\,\,m_k>0,
\end{equation}
where $\mathbf{U}$ is built from eigenvectors of the matrix
$\mathbf{A}=\mathbf{MM^*}$ with the spectrum $\{m^2_k\}$, i.e.
$\mathbf{U^+AU}=diag(m^2_k)$. Consistency of the last relation and
Eq.~(\ref{E:5.6}) is evident from the equality
\begin{equation}\label{E:5.7}
 \mathbf{U^+MU^*(U^*)^{-1}M^*U}=diag(m^2_k),\quad m^*_k=m_k.
\end{equation}

The method based on the Takagi theorem was considered in
Refs.~\cite{12}, \cite{12c}, \cite{21} and \cite{22}, where the
standard way of the determination of the spectrum is given. However, there is
no need to solve this complicated problem in the case considered.
Here we show that the spectrum of the matrix
$\mathbf{A}=\mathbf{M}_I\mathbf{M}^+_I$ coincides with the
squared spectrum of the traditional real mass matrix $\mathbf{M}^{'}_0$.
The spectrum of the matrix
$\mathbf{A}=\mathbf{M}_I\mathbf{M}_I^+$ ($\mathbf{M}_I$ is
defined by (\ref{E:5.4})) follows from the solution of the
characteristic equation $\det (\mathbf{A}-\lambda\cdot\mathbf{1})=0$,
\begin{equation}\label{E:5.8}
 \lambda^4-a\lambda^3+b\lambda^2-c\lambda+d=0.
\end{equation}
The coefficients $a,b,c,$ and $d$ in Eq.~(\ref{E:5.8}) are expressed in
terms of the matrix elements of $\mathbf{M}_0$ as follows:
\begin{eqnarray}\nonumber
 a&=&M^2_1+M^2_2+2\mu^2+2M^2_Z;\\
 b&=&M^2_1M^2_2+2M^2_1(\mu^2+M^2_Z\cos^2
     \theta_W)+2M^2_2(\mu^2+\nonumber\\
  &+&M^2_Z\sin^2\theta_W)+2M^2_Z\mu\sin 2\beta(M_1\sin^2
     \theta_W+\nonumber\\
  &+&M_2\cos^2\theta_W)+(\mu^2+M^2_Z)^2;\nonumber\\
 c&=&2\mu^2M^2_1 M^2_2+M^2_1 [(\mu^2+M^2_Z \cos^2 \theta_W)^2+\nonumber\\
  &+&2M_2M^2_Z\mu\cos^2\theta_W\sin 2\beta]
     +M^2_2[(\mu^2+M^2_Z\sin^2\theta_W)^2+\nonumber\\
  &+&2M_1M^2_Z\mu\sin^2\theta_W\sin2\beta]+
     \frac{1}{2}M_1M_2 M^4_Z \sin^2 2\theta_W+\label{E:5.9}\\
  &+&2M^2_Z\mu^3\sin
     2\beta(M_1\sin^2\theta_W+M_2\cos^2\theta_W)+\nonumber\\
  &+&\mu^2M^4_Z\sin^22\beta;\nonumber\\
 d&=&\mu^4 M^2_1M^2_2+M^2_1\mu^2
     M^2_Z(2\mu M_2+M^2_Z\cos^2\theta_W\sin 2\beta)\times\nonumber\\
 &\times&\cos^2\theta_W\sin2\beta
     +M^2_2\mu^2 M^2_Z(2\mu M_1+M^2_Z\sin^2\theta_W\sin
     2\beta)\times\nonumber\\
 &\times&\sin^2\theta_W\sin2\beta
     +\frac{1}{2}M_1M_2\mu^2 M^4_Z\sin^2 2\theta_W\sin^2
     2\beta.\nonumber
\end{eqnarray}
Analogous expressions are given in Ref.~\cite{12}, where the
algorithm of the definition of the spectrum $\lambda_k$ is considered. In the
general case, we can get exact expressions for the roots $\lambda_k$ of
Eq.~(\ref{E:5.8}) in terms of its general algebraic solutions. It is difficult to analyze and compare such expressions, but we can show that the roots of
Eq.~(\ref{E:5.8}) are $\lambda_k=m^2_k$, where $m_k$ is the
conventional neutralino spectrum. To show this, let us write the
characteristic equation, $\det(\mathbf{M}^{'}_0-l\cdot\mathbf{1})=0$,
in the form ($\mathbf{M}^{'}_0$ is the standard real mass matrix \cite{22},
\cite{24})
\begin{eqnarray}\nonumber
&&{}(M_1-l)(M_2-l)(l^2-\mu^2)+M^2_Z(l+\mu\sin 2\beta)\times\\
&&{}\times(M_1\cos^2\theta_W+M_2\sin^2\theta_W-l)=0.\label{E:5.10}
\end{eqnarray}
Then we arrange the even and odd degrees of $l$ on the left and
right hand sides of this equation separately. Squaring the equation and changing $l^2_k\to\lambda_k$, we get the equation
(\ref{E:5.8}) with coefficients (\ref{E:5.9}). Moreover, by direct
calculation we have checked that the mass spectrum appearing as a
result of the diagonalization
\begin{equation}\label{E:5.11}
\mathbf{U^+M_IU^*}=diag(m_k)
\end{equation}
is entirely positive (see Appendix B).

With the help of the Takagi theorem it is possible to illustrate the correct
construction of the positive mass spectrum. However, the above
discussed method is not convenient for calculations and can be used
as the diagonalizability proof only: there is a unitary matrix
with the property (\ref{E:5.6}) or (\ref{E:5.11}). The use of
$\mathbf{M}_I$ gives a convenient tool for the calculation of the
spectrum and states when $M_1, M_2, \mu$ and their differences are
much greater than $M_Z$. The hierarchy of $M_1$, $M_2$ and $\mu$ is
arbitrary, i.e. one can apply the method suggested to the various
scenarios of the neutralino DM. In this case, the diagonalyzing
matrix is quasidiagonal, i.e. $|U_{kk}|\approx 1$ and $|U_{ik}|\ll
1$, $i\ne k$. Then we represent this matrix in the exponential form
which contains six angles and six phases as input parameters. A similar
approach was considered in the general case in Ref.~\cite{12}, where six
angle and ten phase parameters were applied. Here we show that in
the case of the mass matrix (\ref{E:5.4}) it is possible to use six
phases only (see Appendix B):
\begin{equation}\label{E:5.12}
 \mathbf{U}=\begin{pmatrix}
 a_1&\delta_1 e^{-i\phi_1}&\delta_2 e^{-i\phi_2}&\delta_3
 e^{-i\phi_3}\\r_1e^{i\alpha_1}&a_2&\delta_4 e^{-i\phi_4}&\delta_5
 e^{-i\phi_5}\\r_2 e^{i\alpha_2}&r_4 e^{i\alpha_4}&a_3&\delta_6
 e^{-i\phi_6}\\r_3 e^{i\alpha_3}&r_5 e^{i\alpha_5}&r_6
 e^{i\alpha_6}&a_4\end{pmatrix}\,,
\end{equation}
where $\delta_k$ and $\phi_k$ are input angle and phase parameters.
The values $a_{\beta}, r_i$ and $\alpha_k$ are some functions of the
input parameters which are defined by the unitary conditions
$(\mathbf{U^+U})_{ik}=\delta_{ik}$. In our case, $|\delta_k|\ll 1$
and functions $a_{\beta}, r_i$ and $\alpha_k$ are easily determined
by successive approximations (Appendix B). Apparently, the
diagonalization of the real matrix demands the angle parameters
only. Having used the diagonalization conditions
\begin{equation}\label{E:5.13}
\mathbf{U}^T_I\mathbf{M}_I\mathbf{U}_I=\mathbf{M}_d \equiv
diag(m_1,m_2,m_3,m_4)
\end{equation}
the input parameters $\delta_i$ and $\phi_k$ can be determined from
the six independent equations $(\mathbf{M}_d)_{ik}=0,\,\,i>k$. So, there
are six conditions for the real and six ones for the imaginary parts of matrix
elements. Then the masses $m_{\alpha}$ appear as functions of the
defined input parameters. As is shown in Appendix B, the
perturbative calculation scheme can easily be formalized.

The above discussed method of diagonalization is applied to
the case of a mass matrix with complex parameters $M_1 e^{i\psi_1}$,
$M_2 e^{i\psi_2}$ and $\mu e^{i\psi_{\mu}}$ (see Appendix B). In this case, we have to generalize (\ref{E:5.12}) introducing additional phase parameters
according to $a_k\rightarrow a_k e^{i\xi_k}$. Thus, we have the same
quantity of parameters as in \cite{12}. The functions $a_{\beta},
r_k$ and $\alpha_k$ are determined in terms of the input parameters
$\delta_k, \phi_k$ and $\xi_{\beta}$ using the unitary condition
$\mathbf{U}^+\mathbf{U}=\mathbf{1}$. The input parameters are
determined in terms of the mass matrix elements if we use the
diagonalization conditions
$(\mathbf{U}^T\mathbf{M}\mathbf{U})_{ik}=0,\, i\ne k$ and
$Im(\mathbf{U}^T\mathbf{M}\mathbf{U})_{ii}=0$. Note
that in this case the perturbative calculation scheme (as
for the real mass matrix) is in order also. However, the
expressions are more complicated and bulky, so we give the results
in the first approximation only (Appendix B).

Here we represent the mass spectrum and parameters of the matrix
$\mathbf{U}$ defined by (\ref{E:5.12}), up to terms $\sim
m^2_Z/M^2_a,\,a=1,2$ (Appendix B). The neutralino masses are
\begin{eqnarray}
m_{\chi_1}&=&M_1+\frac{M^2_Z \sin^2 \theta_W}{M^2_1-\mu^2}(M_1+\mu \sin
2\beta),\nonumber\\
m_{\chi_2}&=&M_2+\frac{M^2_Z \cos^2\theta_W}{M^2_2-\mu^2}(M_2+\mu\sin 2\beta),\nonumber\\
m_{\chi_3}&=&\mu+\frac{M^2_Z(1-\sin
2\beta)}{2(M_1+\mu)(M_2+\mu)}\times\label{E:5.14}\\
 &\times&(M_1\cos^2 \theta_W+M_2\sin^2\theta_W+\mu),\nonumber\\
m_{\chi_4}&=&\mu-\frac{M^2_Z(1+\sin
2\beta)}{2(M_1-\mu)(M_2-\mu)}\times\nonumber\\
 &\times&(M_1 \cos^2\theta_W+M_2\sin^2\theta_W-\mu).\nonumber
\end{eqnarray}
From Eqs.~(4.14) one can see that $m_3$ and $m_4$ have the
same sign. The validity of the expressions (\ref{E:5.14}) does not depend on
the hierarchy of $M_a$ and $\mu$. So one can use Eqs.~(\ref{E:5.14}) in
various SUSY scenarios.

Another feature of the diagonalization is the presence of a free
parameter in the structure of neutralino states (Appendix B).
Evidently, this free parameter is a remainder of $O_2$\,-\,symmetry
in the pure higgsino limit, and it does not enter into expressions
for the masses (the last assertion is checked by direct calculation
in the second order approximation).

The structure of the neutralino chiral fields follows from the
transformations
\begin{align}\label{E:5.15}
 &\phi_L=\mathbf{U}\chi_L,\,\,\,\phi_R=(\phi_L)^C=\mathbf{U}^{*}\chi_R,\notag\\
 &\chi_L=\mathbf{U}^{-1}\phi_L=\mathbf{U}^+\phi_L,\,\,\,\chi_R=(\chi_L)^C=
  \mathbf{U}^T\phi_R,
\end{align}
where $\mathbf{U}=\mathbf{U}_2\cdot\mathbf{U}_I$ and
$(\phi)^T=(B,W^3,H^0_1,H^0_2)$. With the help of Eq.~(\ref{E:5.15}) for
the non-chiral neutralino field $\chi=\chi_L+\chi_R$, we get
\begin{equation}\label{E:5.16}
\phi=(Re\mathbf{U}-i\gamma_5 Im\mathbf{U})\chi,\,\,\,
\chi=(Re\mathbf{U}^T+i\gamma_5\,Im\mathbf{U}^T)\phi.
\end{equation}
In the first order of mixing (see Appendix B) the structure of
the neutralino fields is defined by the following expressions
\begin{eqnarray}\nonumber
 \chi_{1}&\approx&
 B+\frac{i}{\sqrt{2}}(\frac{m_1}{M_1+\mu}+\frac{m_2}{M_1-\mu})\gamma_5H^0_1+\\
 &+&\frac{i}{\sqrt{2}}(\frac{m_1}{M_1+\mu}-\frac{m_2}{M_1-\mu})\gamma_5H^0_2,\nonumber\\
 \chi_{2}&\approx&
 W^3-\frac{i}{\sqrt{2}}(\frac{m_3}{M_2+\mu}+\frac{m_4}{M_2-\mu})\gamma_5H^0_1-\nonumber\\
 &-&\frac{i}{\sqrt{2}}(\frac{m_3}{M_2+\mu}-\frac{m_4}{M_2-\mu})\gamma_5H^0_2,\label{E:5.17}\\
 \chi_{3}&\approx&
 \frac{im_1}{M_1+\mu}\gamma_5B-\frac{im_3}{M_2+\mu}W^3+
 \frac{1}{\sqrt{2}}H^0_1+\frac{1}{\sqrt{2}}H^0_2,\nonumber\\
 \chi_{4}&\approx& -\frac{m_2}{M_1-\mu}B+\frac{m_4}{M_2-\mu}W^3+
 \frac{i}{\sqrt{2}}\gamma_5H^0_1-\frac{i}{\sqrt{2}}\gamma_5H^0_2.\nonumber
\end{eqnarray}

Thus, the imaginary part of the transformations contains the factor
corresponding to the redefinition $\chi^{'}=i\gamma_5 \chi$ in the
minimal diagonalization procedure. It was checked by direct
calculation up to the second order that the diagonalization
of the real matrix $\mathbf{M}^{'}_I$ with redefinition of the field
with negative mass gives the same results when the free parameter
is equal to zero (see Appendix B).

It is known that in a wide class of SUSY scenarios the values
of $M_{1,2}$ and/or $\mu$ are of the order of TeV and higher, so the
coefficients in (\ref{E:5.17}) are of the order of $10^{-1}$ or less.
Thus, the mixing terms give a contribution to the physical values
$\sim 1\%$, so the expressions (\ref{E:5.17}) can be used in
practical calculations with a good accuracy.

\section{Conclusion}

It is known that the diagonalization of the neutralino mass form by
the orthogonal real matrix leads to the neutralino mass spectrum
with one negative mass. This has to be taken into account in
calculation rules or by a redefinition of the field with negative
mass. An alternative way is the diagonalization by a unitary complex
matrix which leads to the mass spectrum with all positive masses.
Formally, both the ways are equivalent, but the second one is more
convenient, because it does not demand any modification of the
standard calculation rules.

In this work, we have considered the connection between the mass
sign, the relative parity of the neutralino states and the structure of the
boson-neutralino interaction. These features should be considered in the evaluation of the SI and SD contribution to
neutralino-nucleon scattering. The suggested approach directly
illustrates the existence of one free parameter, generated by
the specific symmetry of the $\mu$-term. When this parameter is equal to
zero, both approaches give the same results. This was strictly shown
in our work for the mass spectrum and states up to the second
approximation.

We suggest a simple and convenient way of diagonalization by
a unitary matrix with the exponential parametrization. Having used this
matrix, we get transparent perturbative formalization of the
diagonalization procedure. This method gives simple
expressions, illustrating the neutralino states structure and
the form of the gauge boson-neutralino interaction. These expressions can be used in
most of the SUSY scenarios with the accuracy $\sim 1\%$ or higher.

\section*{Acknowledgments}

We would like to thank O. Teryaev for many discussions and much
correspondence. This work was supported in part by Grants RFBR
06-02-16215 and 07-02-91557.

\section*{Appendix A}

To explicitly show the appearance of the imaginary unit in the
neutralino mass matrix and for consistency we give here the minimal
part of the SUSY Lagrangian and briefly describe the transformation of
the initial SUSY expressions to the ones in terms of
four-dimensional fields. All definitions and calculations are
in the notation of Refs.~\cite{23} and \cite{24}. We consider
the electro-weak part of the MSSM Lagrangian
\begin{equation}\label{E:7.1}
 L=L_G+L_H+L_{Ph}.
\end{equation}
In Eq.(\ref{E:7.1}), the gauge term $L_G$ has the standard form
\begin{equation}\label{E:7.2}
 L_G=\frac{1}{4}\{(W^{\alpha}W_{\alpha})_{\theta\theta}+
     (\bar{W}_{\dot{\alpha}}\bar{W}^{\dot{\alpha}})_{\bar{\theta}\bar{\theta}}
     +(W^{\alpha}_bW_{\alpha}^b)_{\theta\theta}
     +(\bar{W}^b_{\dot{\alpha}}\bar{W}_b^{\dot{\alpha}})_{\bar{\theta}\bar{\theta}}\},
\end{equation}
where $W_{\alpha}$ and $W_{\dot{\alpha}}$ are $U(1)$ gauge
superfields and $W^b_{\alpha}$ and $W^b_{\dot{\alpha}}$ are $SU(2)$
gauge superfields. The Higgs term contains two chiral superfields with
hypercharges $Y_{1,2}=\pm 1$,
\begin{eqnarray}\nonumber
 L_H&=&\{H^+_1\exp{(g_1G_1-g_2G_2)}H_1+\\
    &+&H^+_2\exp{(-g_1G_1-g_2G_2)}H_2\}_{\theta\theta\bar{\theta}\bar{\theta}}.
\label{E:7.3}
\end{eqnarray}
The phenomenological part contains the so-called $\mu$-term and gauge
soft mass terms:
\begin{eqnarray}\nonumber
L_{Ph}&=&\mu[(H_1\epsilon H_2)_{\theta\theta}+(H^+_1\epsilon
H^+_2)_{\bar{\theta}\bar{\theta}}]-\frac{1}{2}M_1(bb+\bar{b}\bar{b})-\\
     &-&\frac{1}{2}M_2(\omega_a\omega_a+\bar{\omega}_a\bar{\omega_a}),
\label{E:7.4}
\end{eqnarray}
where $\epsilon=i\tau_2$. To define the notation of the components we also
present the expressions for the gauge superfields $G_1$ and $G_2$ in
Wess-Zumino gauge:
\begin{align}\label{E:7.5}
 &G_1=\theta\sigma^{\mu}\bar{\theta}\cdot B_{\mu}+i\theta\theta\cdot\bar{\theta}\bar{b}-
      i\bar{\theta}\bar{\theta}\cdot\theta b+\frac{1}{2}\theta\theta\cdot\bar{\theta}\bar{\theta}\cdot
      D_1,\notag\\
 &G^a_2=\theta\sigma^{\mu}\bar{\theta}\cdot W^a_{\mu}+i\theta\theta\cdot\bar{\theta}\bar{\omega}^a-
      i\bar{\theta}\bar{\theta}\cdot\theta \omega^a+\frac{1}{2}\theta\theta\cdot\bar{\theta}\bar{\theta}\cdot
      D^a_2,\\
 &W_{\alpha}=-\frac{1}{4}\bar{D}\bar{D}D_{\alpha}G,\,\,\,\bar{W}_{\dot{\alpha}}=
       -\frac{1}{4}DD\bar{D}_{\dot{\alpha}}G.\notag
\end{align}
The Higgs chiral superfields are
\begin{equation}\label{E:7.6}
 H_1=h_u+\sqrt{2}\theta h_1+\theta\theta\cdot F_1,\,\,\,H_2=h_d+\sqrt{2}\theta h_2+
 \theta\theta\cdot F_2.
\end{equation}
Thus, the particle content is
\begin{eqnarray}\nonumber
 &&{}G_1=(B_{\mu},b),\,G^a_2=(W^a_{\mu},\omega^a),\\
 &&{}H_1=(h_u,h_1),\,H_2=(h_d,h_2),\label{E:7.7}
\end{eqnarray}
where
\begin{equation}\label{E:7.8}
 h_u=\begin{pmatrix}h^0_u\\h^-_u\end{pmatrix},\,h_1=\begin{pmatrix}h^0_1\\h^-_1\end{pmatrix},\,
 h_d=\begin{pmatrix}h^+_d\\h^0_d\end{pmatrix},\,h_2=\begin{pmatrix}h^+_2\\h^0_2\end{pmatrix}.
\end{equation}
In Eqs.(\ref{E:7.7}) and (\ref{E:7.8}) $B_{\mu},W^a_{\mu},h_u,h_d$ are boson fields and $b,\omega^a,h_1,h_2$ are
two-component fermions.

Using the standard method from Eqs. (\ref{E:7.1})-(\ref{E:7.4}) we
get Lagrangians in terms of
two-component fermions. The gauge field Lagrangian is
\begin{align}\label{E:7.9}
 L_G=&-\frac{1}{4}B^{\mu\nu}B_{\mu\nu}+i\bar{b}\bar{\sigma}^{\mu}\partial
      _{\mu}b+\frac{1}{2}D^2_1-\frac{1}{4}W^{\mu\nu}_aW^a_{\mu\nu}\notag\\
     &+i\omega^a\sigma^{\mu}(\partial_{\mu}\bar{\omega}_a+
       g_2\epsilon_{abc}\bar{\omega}_cW^b_{\mu})+\frac{1}{2}D^a_2D_{2a},
\end{align}
where $B_{\mu\nu}=\partial_{\mu}B_{\nu}-\partial_{\nu}B_{\mu}$ and
$W^a_{\mu\nu}=\partial_{\mu}W^a_{\nu}-\partial_{\nu}W^a_{\mu}+g_2\epsilon^{abc}W_{\mu
b}W_{\nu c}$. The Higgs field Lagrangian is
\begin{align}\label{E:7.10}
L_H&=i\bar{h}_1\bar{\sigma}^{\mu}\partial_{\mu}h_1+i\bar{h}_2\bar{\sigma}^{\mu}\partial_{\mu}h_2-\notag\\
   &-\frac{1}{2}g_1B_{\mu}\bar{h}_1\bar{\sigma}^{\mu}h_1+
    \frac{1}{2}g_1B_{\mu}\bar{h}_2\bar{\sigma}^{\mu}h_2+\notag\\
   &+\frac{1}{2}g_2W^a_{\mu}\bar{h}_1\bar{\sigma}^{\mu}\tau_ah_1
    +\frac{1}{2}g_2W^a_{\mu}\bar{h}_2\bar{\sigma}^{\mu}\tau_ah_2+\notag\\
   &+\frac{ig_1}{\sqrt{2}}(h^+_ubh_1-\bar{h}_1\bar{b}h_u)-
    \frac{ig_1}{\sqrt{2}}(h^+_dbh_2-\bar{h}_2\bar{b}h_d)-\\
   &-\frac{ig_2}{\sqrt{2}}(h^+_u\omega h_1-\bar{h}_1\bar{\omega}h_u)-
    \frac{ig_2}{\sqrt{2}}(h^+_d\omega
    h_2-\bar{h}_2\bar{\omega}h_d).\notag
\end{align}
The phenomenological Lagrangian is
\begin{eqnarray}\nonumber
 L_{Ph}&=&-\mu(h_1\epsilon h_2+\bar{h}_1\epsilon\bar{h}_2)-
 \frac{1}{2}M_1(bb+\bar{b}\bar{b})-\\
 &-&\frac{1}{2}M_2
 (\omega_a\omega_2+\bar{\omega}\bar{\omega}).\label{E:7.11}
\end{eqnarray}

In Eqs. (\ref{E:7.9})-(\ref{E:7.11}) all fermion fields are
two-component spinors. The transition to four-component Mayorana spinors
in a Weyl basis is defined by the following relations
\begin{align}\label{E:7.12}
 &\chi_k=\begin{pmatrix}\phi_k\\
  \bar{\phi}_k\end{pmatrix},\,\,\,\phi_k=(b,\omega^a,h_1,h_2);\notag\\
 &\chi_{kL}=\begin{pmatrix}\phi_k\\0\end{pmatrix},\,\,\,\chi_{kR}=
  \begin{pmatrix}0\\
  \bar{\phi}_k\end{pmatrix},\,\,\,\chi^C_L=\chi_R=R\chi;\\
 &L=\frac{1}{2}(1-\gamma_5)=\begin{pmatrix}1&0\\0&0\end{pmatrix},\,\,\,
  R=\frac{1}{2}(1+\gamma_5)=\begin{pmatrix}0&0\\0&1\end{pmatrix}.\notag
\end{align}

With (\ref{E:7.12}) we get
\begin{align}\label{E:7.13}
 &i\bar{b}\bar{\sigma}^{\mu}\partial_{\mu}b=\frac{i}{2}\bar{B}\gamma^{\mu}\partial_{\mu}B,
  \,\,\,bb+\bar{b}\bar{b}=\bar{B}B,\,\,\,B=\begin{pmatrix}b\\ \bar{b}\end{pmatrix};\notag\\
 &i\omega^a\sigma^{\mu}\partial_{\mu}\bar{\omega}_a=\frac{1}{2}\bar{W}^a\gamma^{\mu}\partial_{\mu}W^a
  +Div(W),\notag\\&\omega_a\omega^a+\bar{\omega}_a\bar{\omega}^a=\bar{W}_aW^a,\,\,\,
  W^a=\begin{pmatrix}\omega^a\\ \bar{\omega}^a\end{pmatrix};\notag\\
 &ig_2\omega^a\sigma^{\mu}\epsilon_{abc}\bar{\omega}_cW^b_{\mu}=
  \frac{i}{2}g_2\bar{W}^a\gamma^{\mu}W^c\epsilon_{abc}W^b_{\mu};\notag\\
 &i\bar{h}_1\bar{\sigma}^{\mu}\partial_{\mu}h_1=\frac{i}{2}\bar{H}_1\gamma^{\mu}\partial_{\mu}H_1,\notag\\
 &\bar{h}_1\bar{\sigma}^{\mu}h_1=\bar{H}_1\gamma^{\mu}H_{1L},\,\,\,
  \bar{h}_2\bar{\sigma}^{\mu}h_2=\bar{H}_2\gamma^{\mu}H_{2L};\\
 &\bar{h}_1\bar{\sigma}^{\mu}\tau_a h_1=\bar{H}_1\gamma^{\mu}\tau_a
 H_{1L},\notag\\
 &\bar{h}_2\bar{\sigma}^{\mu}\tau_a h_2=\bar{H}_2\gamma^{\mu}\tau_a H_{2L},\,\,\,
  H_a=\begin{pmatrix}h^a\\ \bar{h}^a\end{pmatrix};\notag\\
 &h^+_ubh_1-\bar{h}_1\bar{b}h_u=h^+_u\bar{B}H_{1L}-\bar{H}_{1L}Bh_u,\notag\\
 &h^+_dbh_2-\bar{h}_2\bar{b}h_d=h^+_d\bar{B}H_{2L}-\bar{H}_{2L}Bh_d;\notag\\
 &h^+_u\omega
 h_1-\bar{h}_1\bar{\omega}h_u=h^+_u\bar{W}H_{1L}-\bar{H}_{1L}Wh_u,\notag\\
 &h^+_d\omega h_2-\bar{h}_2\bar{\omega}h_d=h^+_d\bar{W}H_{2L}-\bar{H}_{2L}Wh_d;\notag\\
 &h_1\epsilon h_2+\bar{h}_1\epsilon\bar{h}_2=\bar{H}_{1R}\epsilon
  H_{2L}+\bar{H}_{1L}\epsilon H_{2R}.\notag
\end{align}
From Eqs. (\ref{E:7.9})-(\ref{E:7.11}) with the help of Eqs.
(\ref{E:7.12}), (\ref{E:7.13}) we obtain Lagrangians in terms of
four-component spinors. The gauge field Lagrangian is
\begin{align}\label{E:7.14}
 L_G=&-\frac{1}{4}B^{\mu\nu}B_{\mu\nu}+\frac{i}{2}\bar{B}\gamma^{\mu}\partial_{\mu}B
      +\frac{1}{2}D^2_1-\frac{1}{4}W^{\mu\nu}_aW^a_{\mu\nu}\notag\\
     &+\frac{i}{2}\bar{W}^a\gamma^{\mu}(\partial_{\mu}W_a+g_2\epsilon_{abc}W^cW^b_{\mu})
      +\frac{1}{2}D^a_2D_{2a}.
\end{align}
The Higgs fermion field Lagrangian is
\begin{align}\label{E:7.15}
 L_H&=\frac{i}{2}\bar{H}_1\gamma^{\mu}(\partial_{\mu}H_1+ig_1B_{\mu}H_{1L}-
      ig_2W^a_{\mu}\tau_aH_{1L})\notag\\
    &+\frac{i}{2}\bar{H}_2\gamma^{\mu}(\partial_{\mu}H_2-ig_1B_{\mu}H_{2L}-
      ig_2W^a_{\mu}\tau_aH_{2L})\notag\\
    &+\frac{ig_1}{\sqrt{2}}(h^+_u\bar{B}H_{1L}-\bar{H}_{1L}Bh_u)-
      \frac{ig_1}{\sqrt{2}}(h^+_d\bar{B}H_{2L}-\bar{H}_{2L}Bh_d)\\
    &-\frac{ig_2}{\sqrt{2}}(h^+_u\bar{W}H_{1L}-\bar{H}_{1L}Wh_u)-
      \frac{ig_2}{\sqrt{2}}(h^+_d\bar{W}H_{2L}-\bar{H}_{2L}Wh_d).\notag
\end{align}
The phenomenological Lagrangian is
\begin{equation}\label{E:7.16}
 L_{Ph}=-\frac{1}{2}M_1\bar{B}B-\frac{1}{2}M_2\bar{W}_aW^a-\mu(\bar{H}_{1R}\epsilon
 H_{2L}+\bar{H}_{1L}\epsilon H_{2R}).
\end{equation}

\section*{Appendix B}

Here we consider a simple and easily formalized method of the complex
mass form diagonalization:
\begin{equation}\label{E:8.1}
 (\bar{\phi}^I_R)^T\mathbf{M}_I\phi^I_L+h.c.=(\bar{\chi}_R)^T\mathbf{U}^T_I\mathbf{M}_I
 \mathbf{U}_I\chi_L+h.c.=m_i\bar{\chi}_i\chi_i.
\end{equation}
In Eq.~(\ref{E:8.1}) $\mathbf{M}_I$ is a symmetric complex matrix,
$\phi^I_{R,L}$ are the chiral components of the initial Majorana
spinor fields arising after intermediate diagonalization
(\ref{E:5.3}), and $\chi$ are the final Majorana fields
(neutralino):
\begin{equation}\label{E:8.2}
 (\phi^I)^T=(B,W^3,\phi^I_3,\phi^I_4),\,\,\,\chi^T=(\chi_1,\chi_2,\chi_3,\chi_4).
\end{equation}
The intermediate states $\phi^I_3$ and $\phi^I_4$ are defined by
Eq.~(3.2). We suggest straightforward diagonalization of the
form (\ref{E:8.1}) by the unitary matrix in the exponential
parametrization. In the general case, the unitary matrix
$\mathbf{U}(n\times n)$ has $2n^2-n^2=n^2$ parameters, where $n^2$
unitary conditions are taken into account. For $n=4$ we have 16
input parameters, six angles and ten phases \cite{12}. However, in the
case of a symmetric mass matrix with real $M_{1,2}$ and $\mu$, this
number of parameters is excessive, so we suggest a unitary matrix
with six angle and six phase input parameters. These 12 parameters can
be defined from 12 independent conditions following from the symmetric
matrix diagonalization
$(\mathbf{U}^T_I\mathbf{M}_I\mathbf{U}_I)_{ik}=0,\,i>k$ (or $i<k$).
It is convenient for the analysis to use $\mathbf{U}_I$ in the
exponential form \cite{25}:
\begin{equation}\label{E:8.3}
\mathbf{U}_I=\begin{pmatrix}
 a_1&\delta_1 e^{-i\phi_1}&\delta_2 e^{-i\phi_2}&\delta_3
 e^{-i\phi_3}\\r_1e^{i\alpha_1}&a_2&\delta_4 e^{-i\phi_4}&\delta_5
 e^{-i\phi_5}\\r_2 e^{i\alpha_2}&r_4 e^{i\alpha_4}&a_3&\delta_6
 e^{-i\phi_6}\\r_3 e^{i\alpha_3}&r_5 e^{i\alpha_5}&r_6
 e^{i\alpha_6}&a_4\end{pmatrix}.
\end{equation}
In Eq.~(\ref{E:8.3}) $\delta_1-\delta_6$  and $\phi_1-\phi_6$ are
the angle and phase parameters, respectively. The quantities $\delta_k$
and $\phi_k$ are the input parameters, while $a_{\beta}, r_k,
\alpha_k$ are some functions of the input parameters which follow
from the unitary condition $\mathbf{U^+U}=\mathbf{1}$:
\begin{align}\label{E:8.4}
&a_1=(1-\delta^2_1-\delta^2_2-\delta^2_3)^{1/2},\,\,\,
 a_2=(1-\delta^2_4-\delta^2_5-r^2_1)^{1/2},\notag\\
&a_3=(1-\delta^2_6-r^2_2-r^2_4)^{1/2},\,\,\,
 a_4=(1-r^2_3-r^2_5-r^2_6)^{1/2},\notag\\
&a_1r_1e^{i\alpha_1}+a_2\delta_1e^{i\phi_1}+
 \delta_2\delta_4e^{i(\phi_2-\phi_4)}+\delta_3\delta_5e^{i(\phi_3-\phi_5)}=0,\notag\\
&a_1r_2e^{i\alpha_2}+\delta_1r_4e^{i(\alpha_4+\phi_1)}+a_3\delta_2e^{i\phi_2}+
 \delta_3\delta_6e^{i(\phi_3-\phi_6)}=0,\\
&a_1r_3e^{i\alpha_3}+\delta_1r_5e^{i(\alpha_5+\phi_1)}+
 \delta_2r_6e^{i(\phi_2+\alpha_6)}+a_4\delta_3e^{i\phi_3}=0,\notag\\
&r_1r_3e^{i(\alpha_3-\alpha_1)}+a_2r_5e^{i\alpha_5}+
 \delta_4r_6e^{i(\phi_4+\alpha_6)}+a_4\delta_5e^{i\phi_5}=0,\notag\\
&r_1r_2e^{i(\alpha_2-\alpha_1)}+a_2r_4e^{i\alpha_4}+a_3\delta_4e^{i\phi_4}+
 \delta_5\delta_6e^{i(\phi_5-\phi_6)}=0,\notag\\
&r_2r_3e^{i(\alpha_3-\alpha_2)}+r_4r_5e^{i(\alpha_5-\alpha_4)}+
 a_3r_6e^{i\alpha_6}+a_4\delta_6e^{i\phi_6}=0.\notag
\end{align}
The Anzats (\ref{E:8.3}) is convenient for approximate calculations in
the case of a quasidiagonal mass matrix, for instance, $\mathbf{M}_I$
defined by Eq.~(\ref{E:5.4}). In the case considered, the absolute values of
the diagonal elements and the differences are much greater than
the off-diagonal ones (the equality of the third and fourth diagonal
elements $\mu$ in $\mathbf{M}_I$ is compensated by off-diagonal
zero). The diagonalyzing matrix has a similar structure, i.e. the input parameters $\delta_k$ in Eq.~(\ref{E:8.4}) are small, $\delta_k\ll 1$ and $a_k\simeq 1$. So, due to the smallness of parameters $\delta_k\sim M_Z/M_a$, where $a=1,2$, one can easily solve the system of equations (\ref{E:8.4}) approximately. The functions
$a_\alpha,r_k$ and $\alpha_k$ are determined from Eqs.~(\ref{E:8.4}), the input parameters $\delta_k$ and $\phi_k$ are defined by the conditions
$(\mathbf{U}^T_I\mathbf{M}_I\mathbf{U}_I)_{ik}=0,\,i>k$. Hence, the
diagonal elements $(\mathbf{U}^T_I\mathbf{M}_I\mathbf{U}_I)_{kk}=m_k$ give the masses
in terms of known quantities.

Finally, we have done the calculations up to the second order
$\sim M^2_Z/M^2_a$ (or $M^2_Z/\mu^2$) inclusively and get the
expressions for the elements of the diagonalizing matrix (\ref{E:8.3}) (the
hierarchy of $M_a$ and $\mu$ is arbitrary). Such an approximation is
reasonable for calculations within a wide class of Split SUSY
models.

The input parameters are
\begin{align}\label{E:8.5}
 &\delta_1e^{-i\phi_1}=\frac{M^2_Z\sin
  2\theta_W}{2(M^2_2-\mu^2)}\frac{M_2+\mu\sin
  2\beta}{M_1-M_2},\notag\\
 &\delta_2e^{-i\phi_2}=i\frac{m_1}{M_1+\mu},\quad
 \delta_3e^{-i\phi_3}=-\frac{m_2}{M_1-\mu},\notag\\
  &\delta_4e^{-i\phi_4}=-i\frac{m_3}{M_2+\mu},\,\,\,\delta_5e^{-i\phi_5}=\frac{m_4}{M_2-\mu},\\
 &\delta_6e^{-i\phi_6}=-\frac{i}{2\mu}\left(\frac{m_1m_2}{M_1-\mu}+\frac{m_3m_4}{M_2-\mu}\right).\notag
\end{align}
The diagonal elements $a_{\beta}$ are
\begin{align}\label{E:8.6}
 &a_1=1-\frac{1}{2}\,\left[\frac{m^2_1}{(M_1+\mu)^2}+\frac{m^2_2}{(M_1-\mu)^2}\,\right],\notag\\
 &a_2=1-\frac{1}{2}\,\left[\frac{m^2_3}{(M_2+\mu)^2}+\frac{m^2_4}{(M_2-\mu)^2}\,\right],\notag\\
 &a_3=1-\frac{1}{2}\,\left[\frac{m^2_1}{(M_1+\mu)^2}+\frac{m^2_3}{(M_2+\mu)^2},\right],\\
 &a_4=1-\frac{1}{2}\,\left[\frac{m^2_2}{(M_1-\mu)^2}+\frac{m^2_4}{(M_2-\mu)^2},\right].\notag
\end{align}
The off-diagonal elements are
\begin{align}\label{E:8.7}
 r_1e^{i\alpha_1}&=\frac{m_1m_3}{(M_1+\mu)(M_2+\mu)}+\frac{m_2m_4}{(M_1-\mu)(M_2-\mu)}-\notag\\
  &-\frac{M^2_Z\sin 2\theta_W(M_2+\mu\sin 2\beta)}{2(M^2_2-\mu^2)(M_1-M_2)}\,,\notag\\
 r_2e^{i\alpha_2}&=\frac{im_1}{M_1+\mu}\,,\,\,\,r_3e^{i\alpha_3}=\frac{m_2}{M_1-\mu}\,,\notag\\
 r_4e^{i\alpha_4}&=-\frac{im_3}{M_2+\mu}\,,\,\,\,r_5e^{i\alpha_5}=-\frac{m_4}{M_2-\mu}\,,\\
 r_6e^{i\alpha_6}&=\frac{m_1m_2}{M^2_1-\mu^2}+\frac{m_3m_4}{M^2_2-\mu^2}-\notag\\
 &-\frac{i}{2\mu}\left(\frac{m_1m_2}{M_1-\mu}+\frac{m_3m_4}{M_2-\mu}\right).\notag
\end{align}

In Eqs.~(\ref{E:8.5} -- \ref{E:8.7}) we give the zero value to the free parameter
$\delta^0_6$ arising in the first order. It was
checked in the first approximation that the existence of the free
parameter $\delta^0_6$ leads to the phase redefinition of the fields
$\chi_3$ and $\chi_4$. Having applied Eqs.~(\ref{E:8.5}) -- (\ref{E:8.7}), we obtain expressions for the neutralino masses (\ref{E:5.14}):
\begin{equation}\label{E:8.8}
 (\mathbf{U}^T_I\mathbf{M}_I\mathbf{U}_I)_{kk}=m_k
\end{equation}
which are positive. We have checked also the unitary condition
$\mathbf{U}^+_I\mathbf{U}_I=\mathbf{1}$.

The structure of the neutralino chiral fields results from the
transformations
\begin{align}\label{E:8.9}
 &\phi_L=\mathbf{U}\chi_L,\,\,\,\phi_R=(\phi_L)^C=\mathbf{U}^{*}\chi_R,\notag\\
 &\chi_L=\mathbf{U}^{-1}\phi_L=\mathbf{U}^+\phi_L,\,\,\,\chi_R=(\chi_L)^C=
  \mathbf{U}^T\phi_R,
\end{align}
where $\mathbf{U}=\mathbf{U}_2\cdot\mathbf{U}_I$, $\mathbf{U}_2$ is
defined by (\ref{E:2.4}) and $(\phi)^T=(B,W^3,H^0_1,H^0_2)$.

To illustrate the relation between the initial and physical fields we give
transformations in the first order of mixing:
\begin{align}\label{E:8.10}
 B_L&\approx\chi_{1L}+\frac{im_1}{M_1+\mu}\chi_{3L}-\frac{m_2}{M_1-\mu}\chi_{4L},\notag\\
 W^3_L&\approx\chi_{2L}-\frac{im_3}{M_2+\mu}\chi_{3L}+\frac{m_4}{M_2-\mu}\chi_{4L},\notag\\
 H^0_{1L}&\approx\frac{i}{\sqrt{2}}(\frac{m_1}{M_1+\mu}+\frac{m_2}{M_1-\mu})\chi_{1L}-\\
 &-\frac{i}{\sqrt{2}}(\frac{m_3}{M_2+\mu}+\frac{m_4}{M_2-\mu})\chi_{2L}
  +\frac{1}{\sqrt{2}}\chi_{3L}+\frac{i}{\sqrt{2}}\chi_{4L},\notag\\
 H^0_{2L}&\approx\frac{i}{\sqrt{2}}(\frac{m_1}{M_1+\mu}-\frac{m_2}{M_1-\mu})\chi_{1L}-\notag\\
 &-\frac{i}{\sqrt{2}}(\frac{m_3}{M_2+\mu}-\frac{m_4}{M_2-\mu})\chi_{2L}
  +\frac{1}{\sqrt{2}}\chi_{3L}-\frac{i}{\sqrt{2}}\chi_{4L}.\notag
\end{align}
Transformation of the $R$-component can easily be found from the relation
$\chi_R=(\chi_L)^C$. Inverse transformations illustrate the
neutralino structure:
\begin{align}\label{E:8.11}
 \chi_{1L}&\approx
 B_L-\frac{i}{\sqrt{2}}(\frac{m_1}{M_1+\mu}+\frac{m_2}{M_1-\mu})H^0_{1L}-\notag\\
  &-\frac{i}{\sqrt{2}}(\frac{m_1}{M_1+\mu}-\frac{m_2}{M_1-\mu})H^0_{2L},\notag\\
 \chi_{2L}&\approx
 W^3_L+\frac{i}{\sqrt{2}}(\frac{m_3}{M_2+\mu}+\frac{m_4}{M_2-\mu})H^0_{1L}+\notag\\
  &+\frac{i}{\sqrt{2}}(\frac{m_3}{M_2+\mu}-\frac{m_4}{M_2-\mu})H^0_{2L},\\
 \chi_{3L}&\approx -\frac{im_1}{M_1+\mu}B_L+\frac{im_3}{M_2+\mu}W^3_L+
  \frac{1}{\sqrt{2}}H^0_{1L}+\frac{1}{\sqrt{2}}H^0_{2L},\notag\\
 \chi_{4L}&\approx -\frac{m_2}{M_1-\mu}B_L+\frac{m_4}{M_2-\mu}W^3_L-
  \frac{i}{\sqrt{2}}H^0_{1L}+\frac{i}{\sqrt{2}}H^0_{2L}.\notag
\end{align}
The transformation of the non-chiral Majorana field
$\chi=\chi_L+\chi_R$ is
\begin{equation}\label{E:8.12}
\phi=(Re\mathbf{U}-i\gamma_5 Im\mathbf{U})\chi,\,\,\,
\chi=(Re\mathbf{U}^T+i\gamma_5\,Im\mathbf{U}^T)\phi.
\end{equation}
In the first order of mixing from (\ref{E:8.12}) it follows that
\begin{align}\label{E:8.13}
 B&\approx\chi_1-\frac{im_1}{M_1+\mu}\gamma_5\chi_3-\frac{m_2}{M_1-\mu}\chi_4,\notag\\
 W^3&\approx\chi_2+\frac{im_3}{M_2+\mu}\gamma_5\chi_3+\frac{m_4}{M_2-\mu}\chi_4,\notag\\
 H^0_1&\approx
 -\frac{i}{\sqrt{2}}(\frac{m_1}{M_1+\mu}+\frac{m_2}{M_1-\mu})\gamma_5\chi_1+\\
  &+\frac{i}{\sqrt{2}}(\frac{m_3}{M_2+\mu}+\frac{m_4}{M_2-\mu})\gamma_5\chi_2
  +\frac{1}{\sqrt{2}}\chi_3-\frac{i}{\sqrt{2}}\gamma_5\chi_4,\notag\\
 H^0_2&\approx-\frac{i}{\sqrt{2}}(\frac{m_1}{M_1+\mu}-\frac{m_2}{M_1-\mu})\gamma_5\chi_1+\notag\\
  &+\frac{i}{\sqrt{2}}(\frac{m_3}{M_2+\mu}-\frac{m_4}{M_2-\mu})\gamma_5\chi_2
  +\frac{1}{\sqrt{2}}\chi_3+\frac{i}{\sqrt{2}}\gamma_5\chi_4.\notag
\end{align}
The structure of non-chiral neutralino states is
\begin{align}\label{E:8.14}
 \chi_{1}&\approx
 B+\frac{i}{\sqrt{2}}(\frac{m_1}{M_1+\mu}+\frac{m_2}{M_1-\mu})\gamma_5H^0_1+\notag\\
  &+\frac{i}{\sqrt{2}}(\frac{m_1}{M_1+\mu}-\frac{m_2}{M_1-\mu})\gamma_5H^0_2,\notag\\
 \chi_{2}&\approx
 W^3-\frac{i}{\sqrt{2}}(\frac{m_3}{M_2+\mu}+\frac{m_4}{M_2-\mu})\gamma_5H^0_1-\notag\\
  &-\frac{i}{\sqrt{2}}(\frac{m_3}{M_2+\mu}-\frac{m_4}{M_2-\mu})\gamma_5H^0_2,\\
 \chi_{3}&\approx +\frac{im_1}{M_1+\mu}\gamma_5B-\frac{im_3}{M_2+\mu}W^3+
  \frac{1}{\sqrt{2}}H^0_1+\frac{1}{\sqrt{2}}H^0_2,\notag\\
 \chi_{4}&\approx -\frac{m_2}{M_1-\mu}B+\frac{m_4}{M_2-\mu}W^3+
  \frac{i}{\sqrt{2}}\gamma_5H^0_1-\frac{i}{\sqrt{2}}\gamma_5H^0_2.\notag
\end{align}
Thus, the imaginary part of the transformations contains the factor
corresponding to the redefinition $\chi^{'}=i\gamma_5 \chi$ in the
intermediate diagonalization procedure. By direct calculation (up to the
second order) it was checked that the diagonalization of the
real matrix $\mathbf{M}^{'}_I$ with redefinition of the final field with
negative mass gives the same results as in our case with $\delta^0_6=0$.
Note that our formulae (\ref{E:8.10} -- \ref{E:8.14}) coincide with
the corresponding ones from Ref.~\cite{12} if we redefine the neutralino
state with negative mass as $\chi\to i\gamma_5 \chi$.

Now we generalize the method of diagonalization for the case of a mass
matrix with complex parameters $M_1 e^{i\psi_1}$, $M_2 e^{i\psi_2}$
and $\mu e^{i\psi_{\mu}}$. Then we have to extend (\ref{E:8.3})
introducing additional phase parameters according to
$a_{\beta}\rightarrow a_{\beta} e^{i\xi_k}$. The functions
$a_{\beta}, r_k$ and $\alpha_k$ are determined in terms of the input
parameters $\delta_k, \phi_k$ and $\xi_{\beta}$ by the unitary condition
$\mathbf{U}^+\mathbf{U}=\mathbf{1}$. We represent them in the form
\begin{align}\label{E:8.15}
&a_1=(1-\delta^2_1-\delta^2_2-\delta^2_3)^{1/2},\,\,\,
 a_2=(1-\delta^2_1-\delta^2_4-r^2_5)^{1/2},\notag\\
&a_3=(1-\delta^2_2-r^2_4-r^2_6)^{1/2},\,\,\,
 a_4=(1-r^2_3-r^2_5-r^2_6)^{1/2},\notag\\
&a_1r_1e^{i(\alpha_1+\xi_1)}+a_2\delta_1e^{i(\phi_1-\xi_2)}+
 \delta_2\delta_4e^{i(\phi_2-\phi_4)}+\notag\\
&+\delta_3\delta_5e^{i(\phi_3-\phi_5)}=0,\notag\\
&a_1r_2e^{i(\alpha_2+\xi_1)}+\delta_1r_4e^{i(\alpha_4+\phi_1)}
+a_3\delta_2e^{i(\phi_2-\xi_3)}+\notag\\
&+\delta_3\delta_6e^{i(\phi_3-\phi_6)}=0,\\
&a_1r_3e^{i(\alpha_3+\xi_1)}+\delta_1r_5e^{i(\alpha_5+\phi_1)}+
\delta_2r_6e^{i(\phi_2+\alpha_6)}+\notag\\
&+a_4\delta_3e^{i(\phi_3-\xi_4)}=0,\notag\\
&r_1r_2e^{i(\alpha_2-\alpha_1)}+a_2r_4e^{i(\alpha_4+\xi_2)}+
a_3\delta_4e^{i(\phi_4-\xi_3)}+\notag\\
&+\delta_5\delta_6e^{i(\phi_5-\phi_6)}=0,\notag\\
&r_1r_3e^{i(\alpha_3-\alpha_1)}+a_2r_5e^{i(\alpha_5+\xi_2)}+
\delta_4r_6e^{i(\phi_4+\alpha_6)}+\notag\\
&+a_4\delta_5e^{i(\phi_5-\xi_4)}=0,\notag\\
&r_2r_3e^{i(\alpha_3-\alpha_2)}+r_4r_5e^{i(\alpha_5-\alpha_4)}+
a_3r_6e^{i(\alpha_6+\xi_2)}+\notag\\
&+a_4\delta_6e^{i(\phi_6-\xi_4)}=0.\notag
\end{align}
The set of input parameters $\delta_k, \phi_k$ and $\xi_{\beta}$ is
determined in terms of the mass matrix elements utilizing the
diagonalization conditions
$(\mathbf{U}^T\mathbf{M}\mathbf{U})_{ik}=0,\, i\ne k$ and
$Im(\mathbf{U}^T\mathbf{M}\mathbf{U})_{ii}=0$.

In the first approximation from the second condition we get
$\xi_1=\psi_1/2, \xi_2=\psi_2/2$ and $\xi_3=\xi_4=\psi_{\mu}/2$.
From the first condition in the same approximation we get
\begin{align}\label{E:8.16}
 &\delta_1=0\,\,\, or\, \phi_1=\psi_1/2, M_1=M_2;\notag\\
 &\delta_2=\frac{-m_1}{M^2_1-\mu^2}[M^2_1+\mu^2-2\mu M_1
  \cos(\psi_1+\psi_{\mu})]^{1/2},\notag\\
 &\tan(\phi_2-\frac{\psi_1}{2})=-\frac{M_1-\mu}{M_1+\mu}\,\,\cot(\frac{\psi_1+\psi_{\mu}}{2});\notag\\
 &\delta_3=\frac{-m_2}{M^2_1-\mu^2}[M^2_1+\mu^2+2\mu M_1
  \cos(\psi_1+\psi_{\mu})]^{1/2},\notag\\
 &\tan(\phi_3-\frac{\psi_1}{2})=\frac{M_1-\mu}{M_1+\mu}\,\,\tan(\frac{\psi_1+\psi_{\mu}}{2});\\
 &\delta_4=\frac{m_3}{M^2_2-\mu^2}[M^2_2+\mu^2-2\mu M_2
  \cos(\psi_2+\psi_{\mu})]^{1/2},\notag\\
 &\tan(\phi_4-\frac{\psi_2}{2})=-\frac{M_2-\mu}{M_2+\mu}\,\,\cot(\frac{\psi_2+\psi_{\mu}}{2});\notag\\
 &\delta_5=\frac{m_4}{M^2_2-\mu^2}[M^2_2+\mu^2+2\mu M_2
  \cos(\psi_2+\psi_{\mu})]^{1/2},\notag\\
 &\tan(\phi_5-\frac{\psi_2}{2})=\frac{M_2-\mu}{M_2+\mu}\,\,\tan(\frac{\psi_2+\psi_{\mu}}{2});\notag\\
 &\delta_6=0\,\,or\,\,\phi_6=(\psi_2+\psi_{\mu})/4.\notag
\end{align}
All expressions for the input phases can be represented as explicit
functions, for example
\begin{equation}\label{E:8.17}
 \phi_2=\frac{\psi_1}{2}-\arctan(\frac{M_1-\mu}{M_1+\mu}\,\cot(\frac{\psi_1+\psi_{\mu}}{2})).
\end{equation}
From Eqs. (\ref{E:8.16}) one can see that taking account of the
complex degrees of freedom in the mass matrix complicates
the calculations. However, the perturbation scheme of the method is
retained and can easily be formalized. It has to be noted also that
we use the same number of phases as in the general method \cite{12},
and the question of optimization of this number for a special kind
of mass matrix is practically important.

\end{document}